\documentstyle[aps,prl,twocolumn,psfig]{revtex}

\newcommand{\com}[1]{ \: \mbox{#1} \quad }

\begin{document}

\draft

\preprint{not yet submitted}

\title{Periodic Orbit theory for Resonant Tunneling Diodes : comparison with
quantum and experimental results}

\author{D. S. Saraga$^1$,  T. S. Monteiro$^1$ and D. C. Rouben$^2$  }
\address{  \,$^1$Dept. of Physics and Astronomy, University College,
University of  London,
Gower St, London WC1E 6BT, U.K.\\
$^2$Centre for Nonlinear Studies and Department of Physics,
 Hong Kong Baptist University, Hong Kong, P.R. China
}
\date{\today}
\maketitle
\begin{abstract}
We investigate whether the quantal and experimental
amplitudes of current oscillations
of Resonant Tunneling Diodes in tilted fields are obtainable
from Periodic Orbit (PO) theories 
by considering recently proposed PO approaches.  
We show, for the first time, that
accurate 
 amplitudes and frequency shifts for the current oscillations
 (typically to within a few $\%$) 
 can be obtained from a simple analytical formula
 both in the stable (torus-quantization) limit and the unstable 
 regimes of the experiments  
 which are dominated by isolated PO's.  
But we find that the PO approach does not describe quantitatively
the dynamically interesting intermediate experimental regimes 
which appear to be 
dominated by contributions from complex orbits and multiple non-isolated PO's.
We conclude that these regimes will not  easily be described by the usual PO
approach, even with simple normal forms.

\end{abstract}


Mesoscopic systems have been extensively investigated as probes of
 'quantum chaos' in real systems.  Among these, the Resonant Tunneling
 Diode (RTD) in tilted fields, first introduced in 1994 ~\cite{F94} , has
 attracted much attention because of the diversity of observed effects
 it exhibits which have been attributed to Periodic Orbits and 
 'soft chaos'.  For instance, it has
 been used as an experimental probe of spectral fluctuations due to
 unstable periodic orbits ~\cite{F94}, quantum scarring
 ~\cite{W96,NS98a,MDC97}, bifurcations \cite{SS95,Boeb95,TSM96,NS97}, 'ghosts'
 \cite{TD97,SM98} and the torus quantization regime~\cite{SM98}.  All these
 effects manifest themselves through observed oscillations in the
 tunneling current of varying amplitude and frequency.

The well-known Gutzwiller Trace Formula (GTF) \cite{Gutz} is a powerful
 tool in the quantization of chaotic systems. 
It relates the frequencies and amplitudes of oscillations in the {\em density of states} to the actions and
 stabilities of classical periodic orbits (PO's) in a simple analytic
 formula. 
However to date we have no equivalent formula which describes
 the corresponding oscillations in the {\em tunneling current}. 
Hence  some interpretations of the experiments, for instance whether bifurcating  orbits are seen or not, remain controversial ~\cite{F97,NF97,MDC97}.
In ~\cite{NS98} a semiclassical treatment
 of the current was presented, evaluating the Bardeen tunneling matrix element
 numerically within the Wigner phase-space representation. 
Thus they could reproduce some interesting
 experimental features such as regimes of period-doubling of the
 current.

We show here that a simple analytical formula developed by Bogomolny
and Rouben ~\cite{Bog2} can yield quantitatively the $\em{amplitudes}$ of contributions from
isolated periodic orbits - stable or unstable.
This represents the first demonstration that 
the amplitudes 
of the current oscillations may be quantitatively described
by a periodic orbit expansion. 
The formula is in good agreement with results obtained using the approach 
of ~\cite{NS98}. 
The latter approach  provides more flexibility 
 allowing, for instance,  excited initial states (which are not required in our calculation). 
 But the advantage of the new formula lies in its simplicity -it is as
easy to evaluate as the GTF- and in the physical insight which the
analytical expression provides. 
For instance, it exposes a {\em shift}
between the frequency of the Gutzwiller density of states oscillations and
the current oscillations. 
This shift is small ($<1\%$) but the resolution
of the quantum scaled calculations easily exceeds this. 
The comparison
carried out here confirms that this shift is indeed easily detectable and
accurately predicted in terms of classical quantities. 
Also, the formula reproduces observed features due to torus-quantisation, in agreement with quantal calculations and the simple model we proposed previously ~\cite{SM98}.

We have applied here a rigorous test to this formula, since we
have compared it with a broad range of accurate amplitudes obtained from
a scaled quantum spectrum ~\cite{TD97,SM98}. 
We also compare with
the extensive set of experimental data obtained at Bell Labs \cite{Boeb95}.  

Many of the most interesting experimental features, such as period-doublings
occur in an intermediate regime characterized by contributions from
multiple non-isolated orbits and complex PO's (ghosts). 
We find here that
agreement in this regime is qualitative: both the formula
and the approach of ~\cite{NS98} yield the rough range of period-doubling
regions, but the amplitudes are in poor agreement with quantal results and 
experiment. 
In particular, in the two 'ghost' regions we identify, we cannot account 
for the amplitude of the current oscillations, even with normal form
corrections.
The strength and persistence of these contributions remains one of
the most puzzling features of these experiments since in general 
'ghosts' are strongly damped away from the tangent bifurcation
where they appear.

We recall briefly the RTD model ~\cite{F94}. 
In essence, the
physical picture is as follows:  an electric field $F$ (along $x$)
and a magnetic field $B$ in the $x-z$ plane (at tilt angle $\theta$ to the $x$ axis) are applied to a double barrier quantum well. 
Electrons in a 2DEG
accumulate at the first barrier and tunnel through both barriers
giving rise to a tunneling current $I$. 
In the process they probe the classical trajectories -regular or chaotic- 
arising from specular reflection at the barrier walls. 
The current oscillates as a function of applied voltage $V$.
After rescaling with respect to $B$, the dynamics at given $\theta$
and ratio of injection energy to voltage ($R=E/V \sim 0.15$ for the
Bell Lab experiments) depends only on the parameter
$\epsilon=V/LB^2$. 
Regular behaviour occurs at high 
$\epsilon \sim 20000$ (in atomic units), chaotic behaviour
for low $\epsilon \sim 1000$. 
The well width is $L=1200\AA$.  
As we study
here small angles ($\theta \leq 27^\circ$), we neglect the shift
$\delta z \sim d \tan \theta$ due to the mean distance $d$ between
the 2DEG and the left inner barrier ($x=0$). 
At $\theta=0^\circ$ 
the current consists of pure period one oscillations, of amplitude
independent of $\epsilon$, associated with
a straight line PO ($t_0$), bouncing alternately between walls. 
This is our reference current amplitude $I_0$ and we normalise all
amplitudes (semiclassical, quantal and experimental) to $I_0$.

\begin{figure}[htb]
\centerline{\psfig{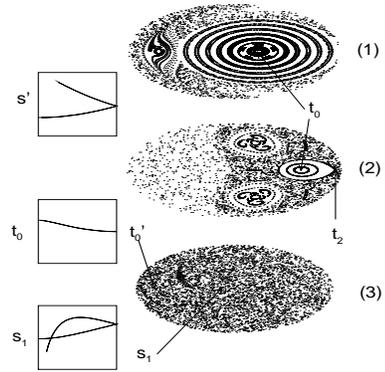}}
\vskip 0.5cm \caption {Shape in $x-z$ plane of the main PO's 
$t_0$ (period-one), $S'$ and $S_1$ (period-2). 
The SOS's illustrate, for $\theta=11^\circ$, three generic
 dynamical regimes typical of $\theta=10-30^\circ$. 
(1) $\epsilon=20000$. Large stable island. 
(2) $\epsilon=7000$. Intermediate regime. $t_0$ about to undergo 
a tangent bifurcation near the edge of the SOS, which will remove the real PO. 
This regime is characterised by contributions from 
non-isolated POs or complex orbits. 
(3) $\epsilon=3000$. Unstable regime. The new $t_0'$ PO has re-appeared
on the far side of the SOS. 
A strong period-two signal from  the isolated unstable PO $S_1$, 
which occupies the central region
most accessible to the tunneling electrons, is seen in the experiments.}
\label{SOS}
\end{figure}

As $\theta$ increases $t_0$ is no longer a straight line
but continues to dominate the period-one oscillations. 
Due to the relatively short coherence time $\tau \sim 0.1$ps 
 ~\cite{F94} just four of the
 shortest PO's ($t_0$, its second traversal $2t_0$ and the
period-two PO's $S_1$ and $S'$) account for all experimental features studied here.
Their shape is shown in Fig. ~\ref{SOS}. 
Nevertheless, their dynamical
behaviour is far from simple. Already in ~\cite{F94} it was observed
that these classical PO's appear and disappear abruptly. 
$t_0$ undergoes a series of tangent bifurcations, where it ceases to exist as
a real PO but leaves a complex 'ghost'. 
Subsequently, below the tangent bifurcation, a new similar looking PO $t_0'$ 
 re-appears from the opposite side of the Surface of Section (SOS) and re-stabilizes .
$S'$ appears abruptly at the discontinuity in the
potential between a barrier and the energy surface due to the magnetic confinement (a 'cusp' bifurcation ~\cite{NS97}). 
It disappears subsequently in a tangent bifurcation also leaving a 'ghost'.

Despite these intricacies, we can identify in the experiments
 three generic dynamical regimes. These are illustrated in the Surfaces of Section in Fig. ~\ref{SOS}. 
They are: 
(1) $\epsilon \gtrsim  20000$. 
The torus regime: the large stable island of $t_0$ yields a period-one
 current. 
The current shows 'jumps' associated with torus quantization.
(2) $20000 \gtrsim \epsilon \gtrsim 3000$. Intermediate regime, with contributions 
from multiple non-isolated PO's or complex orbits. 
(3) $\epsilon \sim 3000-1000$. Unstable regime. 
We can identify contributions from unstable, 
isolated PO's such as $S_1$ (the period-two oscillation
identified in the original Nottingham experiments \cite{F94}). 
We show below that the new semiclassical formula gives excellent results 
in (1) and (3) but rather poor results in the intermediate regime.

The theoretical scaled current (neglecting experimental broadening due to
incoherent processes) is a density of states weighted
by a tunneling matrix element:  $ I(N) =\sum_i{W_i \delta(N-N_i)}$. 
The experimental range ($V=0.1-1.1$ V) corresponds to $N \sim 12-43$, which 
gives an average $N \sim  \hbar^{-1} \sim 28$ corresponding to $V=0.5$ V. 
In fact $N$ is a re-scaled magnetic field ~\cite{TD97,SM98}:
$N=BL \sqrt{2mL\epsilon(R+1/2)}/\pi$. 
In our calculations we used the Bardeen matrix element
\cite{Bard61} form for the tunneling probability. 
Then as explained in ~\cite{Bog2},
one can re-express the matrix element in terms of energy Green's functions
and use their semiclassical expansion
over classical paths.  We consider the initial state describing the
electrons prior of tunneling to be the lowest Landau state:
$\phi_0(z)=\sqrt{B \cos \theta /\pi} \exp(-B \cos \theta z^2/2)$.  Then the
tunneling current is given by:
\[
I(B) \propto  \Re e  \int \! dz \! \! \int \! dz' \sum_{{\rm cl} (z \to z')}
 m_{12}^{-1/2} e^{i S(z,z')} 
e^{-B \cos{\theta} (z^2 +z'^2)/2}
\] 
where $m_{12}=\frac{\partial z}{\partial p_{z_0}}$.

The integrals were evaluated analytically by stationary phase with
 the condition $\partial S/ \partial z=\partial S/\partial z'=0$. 
This condition implies that only 
PO's starting with null momentum $p_z=0$ contribute. 
The resulting contribution to the normalized current $I_{norm}(B)=I(B)/I_0$ 
for a given periodic orbit is approximated by:
\begin{eqnarray}
I_{norm}(B) &=& \Re e \frac{ e^{i B (\tilde{S}+ \Delta \tilde{S}) + i \mu \pi/2 -B \xi}}
{\sqrt{-\cos\theta m_{12} + m_{21}/\cos\theta + 2 i m_{11}}} 
\label{semi}  \nonumber \\
\Delta \tilde{S} &=& \cos\theta z_0^2\gamma/(1+\gamma^2) \com{;} 
\xi=cos\theta z_0^2 \gamma^2 /(1+\gamma^2) \rm \nonumber \\
\gamma &=& \frac{m_{11}-1}{\cos\theta m_{12}} \nonumber \com{,}
\end{eqnarray}
where $\mu$ is a Maslov index, 
$\tilde{S}$ and $m_{ij}$ are the {\em scaled} action and
element of the classical monodromy matrix of the PO, with starting position ($x=0$, $z=z_0$). 
The semiclassical theory predicts that the frequency of the current
oscillations is shifted relative to the scaled action $\tilde{S}$
by $\Delta \tilde{S}$. 

\begin{figure}[ht]
\centerline{\psfig{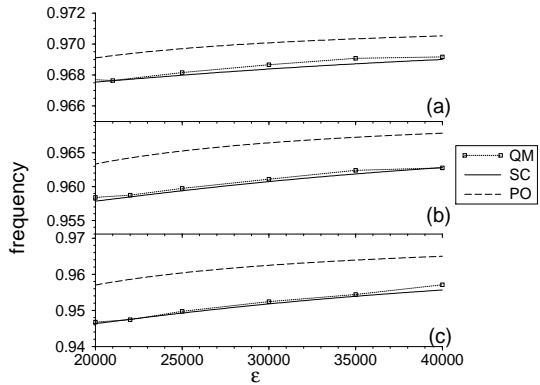}}
\vskip 0.5cm \caption{Frequency of period one oscillation at
(a) $\theta=11^\circ$, (b)$ 20^\circ$ and (c) $27^\circ$. 
Squares = quantum frequency (QM). 
Dashed line = action of the $t_0$ (PO). 
Solid line = semiclassical frequency (SC), which
is shifted relative to the action of the classical PO but is in
 excellent agreement with the quantum results.}
\label{27-action}
\end{figure}

We show in Fig.  \ref{27-action} the scaled action of $t_0$, the
semiclassical frequency $\tilde{S}+ \Delta \tilde{S}$ and the quantum
frequency obtained by Fourier transform.  The shift of frequency is 
$ \sim 1\%$ at most, but clearly we can see that the shifted
frequency is in excellent agreement with the quantal results.  

In \cite{SM98} we explained how one may extract experimental
PO amplitudes in the case where the current has just
a pure period one or two oscillation, by removing
the smooth non-oscillatory component. 
This is only possible in a restricted range of the experiments. 
Also, to compare with theory we must 
in addition consider two factors: 
$i)$ the experimental features are displaced to a lower voltage relative 
to the theoretical value $V=FL$; 
$ii)$ incoherent processes damp each PO contribution by a factor
 $ e^{-T/ \tau}$ where T is the period of the PO. 
For the voltage displacement, we found that all PO features appearing 
at $V=0.5$ V in the calculated spectra are systematically displaced to
a voltage $30\%$ smaller in the experiment. 
For example in \cite{TD97}, 
where characteristic line profiles were correlated with 
different dynamical regimes,
we showed that the distinctive spectral signature of a bifurcation, seen in the quantal spectra  at $\epsilon=13000$,
is seen in the experiment at $\epsilon=10000$. 
Much of this voltage shift is
accounted for by the voltage dependence of the effective mass \cite{SM98}. 
Hence we took $\epsilon \to 1.3 \times \epsilon$ for the experiments, for all angles and all values of $\epsilon$. 
Then we found that, for all angles $\theta=11-27^\circ$,
the position of the theoretical and experimental period-doubling maxima 
are in good agreement.

For the damping we find that the $amplitudes$ of the maximum period-doubling
current would coincide if we chose $\tau$ in the range 
$0.10-0.12$ ps for a given angle. 
This is remarkably consistent with the expected $\tau \sim 0.1$ ps
suggested in \cite{F94}. 
We chose a representative value $\tau=0.11$ ps for all 
the experimental amplitudes. 
In effect period-two amplitudes are damped by incoherent processes by about 
an order of magnitude relative to $I_0$. 

\begin{figure}[ht]
\centerline{\psfig{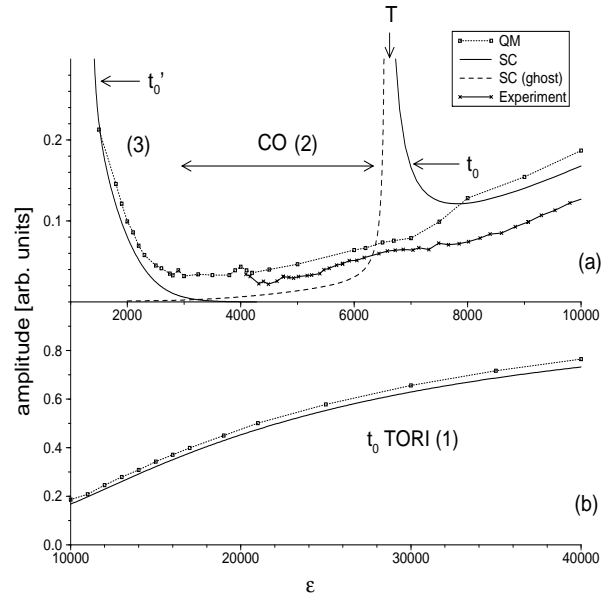}}
\vskip 0.5cm
\caption {Quantal (QM), semiclassical (SC) and experimental amplitudes 
 for period one oscillations at $\theta=11^\circ$ for 
(a) regimes (2) and (3) with tangent bifurcation (T), complex orbit (CO);
(b) torus regime (1). 
The semiclassical formula  gives excellent results where the PO is isolated 
ie for $\epsilon > 8000$ ($t_0$) and $\epsilon< 2000$ ($t_0'$) but gives poor results 
near the bifurcation and for $\epsilon=6500-3000$ in the CO region, even
including the ghost PO. 
Reliable experimental period one amplitudes are only obtainable at $0.5$ V 
from $\epsilon=4000$ up to the start of a period-doubling 
region at $\epsilon \sim 9000$.}
\label{11-log}
\end{figure}

We show in Fig. \ref{11-log} the amplitudes of the experimental, quantal and
semiclassical period one current at $\theta=11^\circ$ from the torus regime
through to the unstable regime. 
For high $\epsilon$ [regime (1)] the semiclassical current in 
Fig. \ref{11-log}(b) is due to the stable $t_0$ and
its tori, and the agreement with the quantal calculation is excellent. 
However, the most interesting dynamical region is at 
$\epsilon=3000-6000$, where there is a
significant quantal/experimental current but no real $t_0$ PO 
($t_0$ disappears in a tangent bifurcation at $\epsilon=6500$, 
 and reappears at $\epsilon=4300$; between $4300 -3000$ the new $t_0$ is not easily accessible to the tunneling electrons
so its contribution is negligible). 
Here [regime (2)] the semiclassical result is poor even when we include 
the 'ghost' complex PO. 
Even with a cubic normal form, which we do not present here, 
agreement remains poor. 
Quantitative agreement is once again good
when the 're-born' isolated real $t_0'$ orbit dominates the current 
for $\epsilon < 2500$ [regime(3)].

\begin{figure}[htb]
\centerline{\psfig{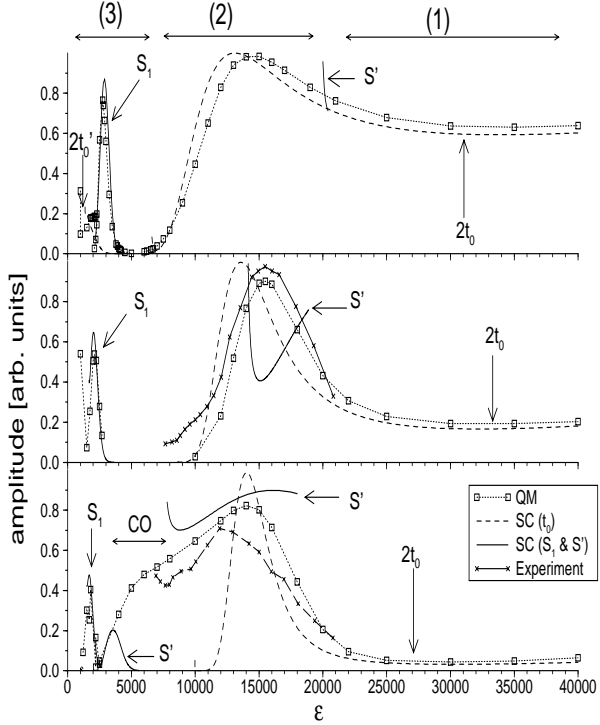}}
\vskip 0.5cm \caption {Quantal (QM), semiclassical (SC) and experimental 
amplitudes for period two current at (a) $\theta=11^\circ$, (b) $\theta= 20^\circ$ and (c) $\theta= 27^\circ$. 
As previously we have 3 regimes. (1) stable 
and (3) unstable isolated PO, where the semiclassical formula is good. 
In regime (2) in contrast we have the non-isolated contributions
of $2t_0$ and $S'$ which cannot be added straightforwardly.
Hence we simply show the individual contributions. 
Here agreement between the formula and quantum calculations/ experiment 
is poor. 
$S'$ appears abruptly at a cusp bifurcation ($\epsilon \sim 18000$)
and disappears in a tangent bifurcation at lower $\epsilon$, below which the experiment shows a slowly
decaying 'plateau' due to a complex orbit (CO).} 
 \label{11-20-27}  
 \end{figure} 

In Fig. \ref{11-20-27} we show a comparison between experimental, quantal
and semiclassical amplitudes of the period two current at $\theta=11^\circ,
20^\circ$ and $27^\circ$.  We were unable to read reliable
experimental period-two amplitudes for $\theta=11^\circ$ as a strong period one beat is also present.
At $\theta=20^\circ$ and $27^\circ$, the quantal calculations and the experiments are in very good agreement. 
As expected in the large stable island regime $\epsilon >25000$
[regime (1)] agreement between the semiclassics and quantum is excellent. 
This is also the case for $\epsilon< 3000$ [regime (3)]. 
Here, the isolated unstable PO $S_1$ 
describes the current very well.

However, in the intermediate regime (2), the quantum current 
requires a coherent
superposition of the non-isolated PO's $2t_0$ and $S'$. 
A straightforward sum (allowing for their phases) yields poor results. 
$2t_0$ and $S'$ have near identical actions, unresolvable in 
the quantum Fourier transform spectroscopy. 
The $2t_0$ contribution comes from Miller tori localized
on the large stable island and, due to the moderate values of $\hbar$,
substantially beyond the island boundary. 
A phase-space analysis with Wigner and Husimi functions
shows that the $S'$ scars are mixed in with the outer tori of the $t_0$
island. 
Hence both their action and phase-space localization
coincide. 
At $11^\circ$ however, the contribution of $S'$ is small
and occupies a narrow range in $\epsilon$. 
In this case the semiclassical amplitudes are quite good. 
This is not the case at $27^\circ$. 
Both the individual island ($2t_0$) and $S'$ contributions are significant 
between $\epsilon =20000-12000$ and there is
no agreement with the quantal results. 
For $\epsilon < 8000$ the $2t_0$
contribution is negligible and $S'$ has disappeared into the complex plane
in a tangent bifurcation. 
Even including the $S'$-ghost, we were unable to obtain quantitative
agreement in this complex orbit (CO) region spanning
$\epsilon=8000-5000$. 
We note that the slowly declining 'plateau' seen quantally and 
in the experiment is a
surprising and unexpected feature, since 'ghosts' should be exponentially
suppressed as the imaginary component of the action grows.

We have also compared with the alternative semiclassical 
approach in ~\cite{NS98}. 
All regimes considered here and in ~\cite{NS98} involved
PO's with initial $p_z=0$. 
In that case, and assuming the lowest Landau state for the initial state,  
we have found that the numerical integral (5) of ~\cite{NS98} reduces to
the analytical formula here.
Both encounter the same difficulties in regime (2). 
We note that although ~\cite{Bog2} requires the $p_z=0$ selection rule,
 ~\cite{NS98} requires only $p_z$ to be {\em small}. 
As yet we have no
unambiguous experimental detection of a PO with $p_z>0$.

Finally, we note that in general, the approach of ~\cite{Bog2} would predict 
complex stationary phase points. 
Their complex part has been  neglected in order that
the theory may obtain PO's.  
Our work \cite{SM99} indicates that the consistent
failure of the PO formalism in the intermediate regime, despite the
usual normal form corrections, may require that
the usual PO picture be partly abandoned since 
complex {\em non-periodic contributions} -as opposed to 'ghosts'
which are complex PO's- may be essential.

We are greatly indebted to E. Bogomolny, E. Narimanov and D. Stone
for helpful discussions. 
We also wish to thank G. Boebinger for providing us his experimental data. 
T. S. M. acknowledges funding from the EPSRC. 
D. S. S. acknowledges the financial support from the TMR programme.

\end{document}